\documentclass[12pt]{article} 
\usepackage{epsf,a4}

\begin{document}
\typeout{! }
\typeout{! }

\newcommand{\D}{\displaystyle}

\newlength{\bi} \setlength{\bi}{15.5cm}
\newcommand{\bild}[4]{\begin{figure}\unitlength1cm
  \addtolength{\bi}{-#2cm}
  \begin{minipage}{0.5\bi}\end{minipage}\hfill\begin{minipage}{#2cm}
    \framebox(#2,#3){\epsfxsize=#2cm \epsfysize=#3cm \epsffile{#1}}
\caption{#4}\end{minipage}\hfill
\begin{minipage}{0.5\bi}\end{minipage}
\end{figure}}

\thispagestyle{empty} \setcounter{page}{0} \null \vfil \vskip 60pt
\begin{center}%
{\LARGE Flow Equations for Electron-Phonon Interactions \par}%
\vskip 3em {\large \lineskip .75em
\begin{tabular}[t]{c} Peter Lenz\footnotemark and Franz Wegner\\[1em]
  Institut f\"ur Theoretische Physik\\ Ruprecht-Karls-Universit\"at\\ 
  D-69120 Heidelberg, Germany
\end{tabular}\par}%
\footnotetext{New address:
    Max-Planck-Institut f\"ur Kolloid- und Grenzfl\"achenforschung,
    Kantstra{\ss}e 55, \mbox{14513 Teltow,} Germany}
\vskip 1.5em
{\large \today \par}%
\end{center}\par
\vskip 3em
\begin{center}
  {\bf Abstract}
\end{center}

{\small A recently \cite{Weg} proposed method of continuous unitary
  transformations is used to eliminate the interaction between
  electrons and phonons. The differential equations for the
  couplings represent an infinitesimal
  formulation of a sequence of Fr\"ohlich-transformations \cite{Fro2}.
  The two approaches are compared. Our result will turn out to be less
  singular than
  Fr\"ohlich's. Furthermore the interaction between  electrons belonging
 to a Cooper-pair will always be attractive in our approach. Even in the case
 where Fr\"ohlich's transformation is not defined (Fr\"ohlich actually 
excluded these regions from the transformation), we obtain an elimination
of the electron-phonon interaction. This is due to a sufficiently slow
change of the phonon energies as a function of the flow parameter.} \newpage
\section{Introduction}

Around 40 years ago Bardeen, Cooper and Schrieffer developed their
famous theory of superconductivity \cite{BCS}. Essential for their
success was the interpretation of an effective interaction between
electrons of a many-particle system \cite{Cop}. This interaction was
assumed to be present in addition to a Coulomb-interaction between the
electrons. As Fr\"ohlich showed 1952 \cite{Fro2} this effective
electron-electron interaction can have its origin in the interaction
between phonons and electrons. In this work Fr\"ohlich eliminated the
electron-phonon interaction by a unitary transformation. By doing so
he was able to describe the interaction mediated by the lattice as an
effective electron-electron interaction.

But Fr\"ohlich's approach contains some problems. He had to exclude 
certain regions in momentum space from the elimination since in these 
regions the transformation would become singular due to a vanishing 
energy-denominator.

In this paper we will apply an elimination procedure recently developed 
by one of the authors \cite{Weg}. Instead of transforming the Hamiltonian in
one step in this new approach the desired transformed Hamiltonian will
be achieved step by step. Or more formally spoken instead of one
unitary transformation a sequence of unitary transformations will be
applied for diagonalization. In an infinitesimal formulation of this
continuous transformation the renormalization of the coupling
constants is described by the flow equations. In order to develop
these differential equations some approximations will be necessary.
By means of these transformations new types of interactions mainly involving 
larger numbers of particles will be generated. They will be neglected after 
normal-ordering. Fr\"ohlich had to use similar approximations.

Our approach has the following advantages: (i) The original phonon-coupling 
can be completely eliminated, even when the states connected by this 
interaction are degenerate. The continuous transformation is chosen in such 
a way that the transformed Hamiltonian does not contain any interactions 
between one electron and the creation or annihilation of one phonon. These 
interactions are still present in
Fr\"ohlich's approach due to normal ordering of generated interactions not 
taken into account. (ii) Singularities in the induced 
electron-electron interaction which appear in Fr\"ohlich's scheme, 
will either not 
appear in our approach or the divergencies will be less singular. In order 
to prove these properties the influence of the
renormalization of the one-particle energies on the elimination of the
electron-phonon coupling has to be taken into account. 
By doing so it is possible to make statements which cannot be obtained
by perturbation theory. Within this approach the way the
couplings reach their renormalized value can be determined. By
discussing the consequences of this asymptotic behavior it becomes
clear that our approach is superior to Fr\"ohlich's.

Our paper is organized as follows. In the following section a short
review of Fr\"ohlich's approach will be given. In the Section \ref{cFe} the
flow equations describing the elimination of the electron-phonon
interaction will be derived. In the next section the above mentioned
approximations made by Fr\"ohlich will be applied to the system of
differential equations. The renormalized values will be calculated.
Also the differences in the electron-electron interactions will be
discussed. Section \ref{sas} contains the analysis of the behavior of our
transformation in the asymptotic regime. The fundamental
differential-equation for the phonon-energies will be discussed. In
the next section the consequences of the asymptotic behavior of the
couplings will be given. It will be shown that the electron-phonon
coupling always is eliminated even in the case of degeneracies. The
last section contains a short summary and outlook.

\section{Fr\"ohlich's Transformation}
\label{sft}
In the following sections we will often refer to the above mentioned
work of Fr\"ohlich \cite{Fro2}. Thus for convenience a short summary
of his results will be given here.

The Hamiltonian of the model will be written as
\begin{eqnarray}
  H & = & \sum_{q}\omega_q:a^{\dagger}_qa_q:+\sum_{k} \varepsilon _k
  :c^{\dagger}_kc_k:+ E+\sum_{k,q}
  M_q(a^{\dagger}_{-q}+a_q)c^{\dagger}_{k+q}c_k \nonumber \\ & \equiv
  & H_0+H_{e-p} .  \label{kT0}
\end{eqnarray}
Here and in the following $k$ stands for $k=\{{\bf k},\sigma\}$, i.e.
the spin is conserved by the electron-phonon interaction thus no
spin-subscript is needed. $a^{(^\dagger)}$ are bosonic creation
respectively annihilation operators. $c^{(\dagger)}$ denote the
corresponding fermionic operators. $:...:$ denotes normal-ordering and
$E$ is a constant energy. Further $M_q$ is the coupling between
electrons and phonons. Following the approach of Bloch \cite{Blo} or
Nordheim \cite{Nor} it is independent of the electron momentum. If
there is need to specify $\varepsilon_k$ or $\omega_q$ a quadratic
dispersion for electrons and a linear dispersion for phonons will be
assumed. Finally it should be emphasized that neither the Coulomb
repulsion nor umklapp-processes will be taken into consideration.

Fr\"ohlich eliminated the electron-phonon interaction in Eq.
(\ref{kT0}) up to order $|M_q(0)|^2$ by expanding a unitary
transformation with the Baker-Hausdorff formula
\[ H^F =  e^{-S}He^{S} =  H+[H,S]+\frac{1}{2}[[H,S],S]+...\]
Fr\"ohlich's ansatz
\begin{equation}
  S:=-\sum_{k,q} M_q \left(\frac{1}{\varepsilon
    _{k+q}-\varepsilon_k+\omega_q}
  a^{\dagger}_{-q}+\frac{1}{\varepsilon _{k+q}-\varepsilon_k -\omega_q
    }a_q \right)c^{\dagger}_{k+q}c_k , \label{kT3}
\end{equation}
assures that the relation
\begin{equation}
  H_{e-p}+[H_0,S]=0
\end{equation}
is fulfilled. If degeneracies can be excluded the transformed
Hamiltonian becomes
\begin{eqnarray}
  H^F & = & \sum_{q} \omega_q^F
  :a^{\dagger}_qa_q:+\sum_{k}(\varepsilon _k^F-2\sum_{\delta}
  n_{k+\delta }V_{k,k+\delta ,\delta }^F) :c^{\dagger}_kc_k: \nonumber
  \\ & & +\sum_{k,k',\delta}
  V^F_{k,k',\delta}:c^{\dagger}_{k+\delta}c^{\dagger}_{k'-\delta}c_{k'}c_k:+
  E^F \nonumber \\ & & + \mbox{ irrelevant Terms } ,\nonumber
\end{eqnarray}
where the irrelevant terms are either of order $|M_q|^3$ or represent
other interactions not taken into account in Eq. (\ref{kT0}). The
occurring coefficients are given by:
\begin{eqnarray}
  \varepsilon _k ^F & = & \varepsilon _k - \sum_{q} |M_q|^2 \left (n_q
  \frac{1}{ \varepsilon _{k+q}-\varepsilon _k-\omega _q}+
  (n_q+1)\frac{1}{\varepsilon _{k+q}-\varepsilon_k+\omega_q}\right) \\ 
  \omega _q ^F & = & \omega _q -\sum_{k} |M_q|^2 \cdot n_k
  \left (\frac{1}{\varepsilon_{k+q}-\varepsilon_k+\omega_q}
  +\frac{1}{\varepsilon_{k+q}-\varepsilon_k-\omega_q}\right)\\ V_{k,k',q} ^F
  & = & |M_q| ^2 \frac{\omega _q}{(\varepsilon _{k+q}-\varepsilon
    _k)^2- \omega _q ^2}\\ E^F & = & E +\sum_{k}n_k
  (\varepsilon_k^F-\varepsilon_k)-\sum_{k,q}n_kn_{k+q}V_{k,k+q,q}^F
\end{eqnarray}
In the last equations the convention has been introduced that $q$
denotes a phonon-wave\-vec\-tor, thus $n_q$ is a bosonic occupation
number whereas $n_k$ and $n_{k+q}$ denote the fermionic ones.

\section{The Flow Equations}
\label{cFe}
In order to eliminate the electron-phonon coupling by flow equations
one divides $H$
\[ H=H^d+H^r  ,\]
into the phonon-number conserving part
\begin{eqnarray*}
  H^d & = & \sum_{q} \omega_q:a^{\dagger}_qa_q:+\sum_{k}
  (\varepsilon_k-2\sum_{\delta} n_{k+\delta }V_{k,k+\delta ,\delta
    }):c^{\dagger}_kc_k: \\ & & +\sum_{k,k',\delta}
  V_{k,k',\delta}:c^{\dagger}_{k+\delta}c^{\dagger}_{k'-\delta}c_{k'}c_k:
  +E\\ & \equiv & H^{ph}+H^{e}+H^{e-e}+E  
\end{eqnarray*}
and the phonon-number violating part
\begin{eqnarray*}
H^r & \equiv &
  H_{e-p} = \sum_{k,q}
  (M_{k,q}a^{\dagger}_{-q}+M^*_{k+q,-q}a_q)c^{\dagger}_{k+q}c_k.
\end{eqnarray*}
All occurring coefficients have to be regarded as functions of the
flow parameter $l$. The initial conditions of the introduced new
quantities are
\begin{eqnarray*}
  M_{k,q}(l=0) & = & M^*_{k+q,-q}(l=0) = M_q(0) \equiv M_q \\ 
  V_{k,k',\delta} (l=0) & = & 0 .
\end{eqnarray*}
The generator of the continuous unitary transformation will be chosen
as
\begin{equation}
  \eta:=\sum_{k,q} (M_{k,q}\alpha _{k,q}a^{\dagger}_{-q}+M^*_{k+q,-q}
  \beta_{k,q}a_q)c^{\dagger}_{k+q}c_k, \label{fg0a}
\end{equation}
where
\[ \alpha_{k,q} = \varepsilon_{k+q}-\varepsilon_k +\omega_q , \quad
\beta_{k,q} = \varepsilon_{k+q}-\varepsilon_k -\omega_q.\]

By calculating the commutator $[\eta ,H]$ one obtains for $dH/dl$ besides 
terms of the type already presented in $H^d$ contributions of the type 
 $W^{2ph}$, $W^{e-2ph}$ and $W^{2e-2ph}$. They are of the form
\begin{eqnarray}
  W^{e-2ph} & \sim & \sum_{k,q,q'}
  (a^{\dagger}_{-q'}a^{\dagger}_{-q}+:a^{\dagger}_{-q'}a_q:+a_{q'}a_q)
  :c^{\dagger}_{k+q+q'}c_k: \label{tH1} \\
  W^{2e-2ph} & \sim & \sum_{k,k',q,q'} 
  (a^{\dagger}_{-q}+a_q):c^{\dagger}_{k+q+q'}c^{\dagger}_{k'-q'}c_{k'}c_k: 
  \label{tH1b} \\
  W^{2ph} & \sim & \sum_{q}
  (a^{\dagger}_{q}a^{\dagger}_{-q} +a_{-q}a_q) . \label{tH2}
\end{eqnarray}
The interactions $W^{e-2ph}$ and $W^{2e-2ph}$ are not included in the 
original model Hamiltonian. They describe two-phonon-processes and the 
interaction of a phonon with two electrons. Although a more realistic 
Hamiltonian should contain such interactions, we will neglect them as 
Fr\"ohlich did.\\ 
The interaction $(\ref{tH2})$ describes the generation and 
annihilation of two phonons. It can be transformed away by introducing
\begin{eqnarray}
  \tilde{H} & = & H+\sum_{q} \mu _q \left (
  a^{\dagger}_qa^{\dagger}_{-q}+ a_qa_{-q}\right ) \nonumber \\ 
  \tilde{\eta} & = & \eta +\eta ^{(2)} , \nonumber \\ & \equiv & \eta
  + \sum_{q} \xi _q \left ( a^{\dagger}_qa^{\dagger}_{-q}
  -a_qa_{-q}\right ) , \nonumber
\end{eqnarray}
where $\mu_q$ and $\xi_q$ are real and $\mu_{q}(l=0)=0$.

These additional terms modify the equations for $\omega_q ,E,M_{k,q}$.
Besides this one obtains the additional flow equation
\[\frac{d\mu _q}{dl}  =  -2\omega _q \xi_q +\sum_{k} n_k\left (M_{k,q}
M_{k+q,-q}\alpha_{k+q,-q}-M_{k-q,q}M_{k,-q}\alpha_{k,-q} \right ). \]
By choosing
\[ \xi_q  =  \frac{1}{2\omega _q} \sum_{k} n_k \left (M_{k,q}M_{k+q,-q}
\alpha_{k+q,-q}-M_{k-q,q}M_{k,-q}\alpha_{k,-q} \right ) \]
it is assured that the interaction (\ref{tH2}) will not be generated,
i.e. one obtains for all flow-parameters $l \in R_0^{+}$
\[ \mu _q (l)=0. \]
Therefore the desired transformed Hamiltonian becomes
\begin{eqnarray}
  H^d (\infty) & = & \sum_{q} \omega_q (\infty)
  :a^{\dagger}_qa_q:+\sum_{k}(\varepsilon _k(\infty) -2\sum_{\delta}
  n_{k+\delta }V_{k,k+\delta ,\delta } (\infty)):c^{\dagger}_kc_k:
  \nonumber \\ & & +\sum_{k,k',\delta}
  V_{k,k',\delta}(\infty):c^{\dagger}_{k+\delta}c^{\dagger}_{k'-\delta}c_{k'}
c_k:+
  E(\infty) \nonumber \\ & & + \mbox{ irrelevant terms } \label{tH3}.
\end{eqnarray}
The renormalization of the coefficients is described by the flow
equations:
\begin{eqnarray}
  \frac{dM_{k,q}}{dl} & = & -\alpha_{k,q}^2M_{k,q}\nonumber \\ 
  & & -2
  \cdot \sum_{\delta} V_{k,k+q+\delta ,\delta }M_{k+\delta
    ,q}\alpha_{k+\delta ,q}\cdot \left( n_{k+q+\delta }-n_{k+\delta }\right
  ) \nonumber \\ 
  & & - 2M_{k,q}\alpha_{k,q}\cdot \sum_{\delta}
  \left(n_{k+\delta }V_{k,k+\delta ,\delta }-n_{k+q+\delta
      }V_{k+q,k+q+\delta ,\delta } \right) \nonumber \\ 
  & &+ 2 \cdot
    \sum_{k'} V_{k,k'+q,q}M_{k',q}\alpha_{k',q} \cdot \left(n_{k'+q}-n_{k'}
  \right) \nonumber \\ 
  & & - \frac{M^*_{k+q,-q}}{\omega _q} \sum_{k'}
  M_{k',q}M_{k'+q,-q}\beta_{k',q} \cdot (n_{k'+q}-n_{k'}) \label{fg1}
  \\ \frac{dV_{k,k',q}}{dl} & = &
  M_{k,q}M^*_{k'-q,q}\beta_{k',-q}-M^*_{k+q,-q}M_{k',-q}\alpha_{k',-q}
  \label{fg3} \\ \frac{d\omega_q}{dl} & = & 2 \cdot \sum_{k}
  |M_{k,q}|^2\alpha_{k,q} \cdot (n_{k+q}-n_k) \label{fg4} \\ 
  \frac{d\varepsilon _k}{dl} & = & -\sum_{q}
  (2n_q|M_{k+q,-q}|^2\beta_{k,q}+2(n_q+1)|M_{k,q}|^2\alpha_{k,q})
\label{fg5} \\
\frac{dE}{dl} & = & \sum_{k} n_k\frac{d\varepsilon _k}{dl} -\sum_{k,q}
n_kn_{k+q}\frac{V_{k,k+q,q}}{dl} . \label{fg6}
\end{eqnarray}
The generated new interactions have been neglected. Thus the last
equation does not describe correctly the renormalization of the ground
state energy. Hence it will not be of further interest.\\ 

\section{Comparison with Fr\"ohlich's Method I: The Non-Degenerate Case}
\label{ssa}

In this first approach it will be assumed that $\alpha_{k,q} \not =0$
holds for all $k,q$. This could be realized by a finite system with
appropriately chosen electron and phonon-dispersion. In this case it is
assured that the Fr\"ohlich-transformation is welldefined.

In this section the flow equations will be compared with Fr\"ohlich's
approach of Section \ref{sft}. His results were only exact up to order
$|M_q(0)|^2$. Thus in the flow equation approach terms of order
$|M_q(0)|^3$ and higher might also be neglected. Because of the
initial condition $V_{k,k',q}(0)=0$ Eq. $(\ref{fg3})$ shows that
$V_{k,k',q}(l)$ is of the order $|M_q(0)|^2$. Thus the lines two to
four in Eq. $(\ref{fg1})$ become irrelevant. The same holds for
the last line of this equation, i.e. is also of the order
$|M_q(0)|^3$. Within the order given by this approximation Eq.
(\ref{fg1}) can be solved exactly
\begin{equation}
  M_{k,q}(l) = M_q(0) e^{- \left ( \varepsilon _{k+q}(0) -\varepsilon
    _k(0)+ \omega _q (0)\right )^2 \cdot l} + {\cal O}(|M_q(0)|^3)
  .\label{fg7e}
\end{equation}

Using this solution the equations for $V_{k,k',q}(l)$, $\varepsilon
_k(l)$ and $\omega _q(l)$ are easily integrated. Thus their
renormalized values can be obtained. While $\varepsilon
_k(\infty)=\varepsilon_k^F$ and $\omega _q (\infty)=\omega _q^F$ holds
the interaction becomes
\begin{equation}
  V_{k,k',q} (\infty) = |M_q(0)| ^2 \left ( \frac{\beta
    _{k',-q}}{\alpha ^2_{k,q}+\beta ^2_{k',-q}} - \frac{\alpha
    _{k',-q}}{\beta ^2_{k,q}+\alpha ^2_{k',-q}} \right) .  \label{sA3}
\end{equation}
Actually the last equation does not represent Fr\"ohlich's result. To
illustrate this a look at the interaction between the
electrons of a Cooper-pair will be taken ($k'=-k$):\\ In this case
Fr\"ohlich proposes (actually independently of this specialization)
\begin{equation}
  V^{F}_{k,k',q} = V^{F}_{k,-k,q} =|M_q(0)|^2\frac{\omega
    _q}{(\varepsilon _{k+q}-\varepsilon_k)^2-\omega _q^2} \quad
  ,\label{sA4}
\end{equation}
whereas the flow equations yield
\begin{equation}
  V_{k,-k,q}(\infty) =- |M_q(0)|^2\frac{\omega _q}{(\varepsilon
    _{k+q}-\varepsilon_k)^2+\omega _q^2} \label{sA5} .
\end{equation}
Thus a remarkable difference between continuous unitary transformations
and the Fr\"oh\-lich transformation has arisen. Before discussing the
origin of this difference it should be mentioned that the difference
between the two interactions is independent of $M_q(0)$, i.e. holds
even for arbitrarily weak coupling.

But it is easily seen that both approaches yield the same result for
{\em real processes}. In this case the one-particle energy is conserved, i.e.
\begin{equation}
  \varepsilon _{k+q}-\varepsilon _k = \varepsilon _{k'}- \varepsilon
  _{k'-q} . \label{sA6}
\end{equation}
For these processes the interaction becomes
\begin{equation}
  V_{k,k',q}^{d} =|M_q(0)|^2\frac{\omega _q}{(\varepsilon
    _{k+q}-\varepsilon_k)^2-\omega _q^2} \mbox{ {\bf iff } }
  \varepsilon _{k+q}-\varepsilon _k = \varepsilon _{k'}- \varepsilon
  _{k'-q} . \label{sA7}
\end{equation}
By introducing
\[ B:=\{ (k,k',q):\varepsilon _{k+q}-\varepsilon _k = \varepsilon _{k'}-
\varepsilon _{k'-q} \}  , \]
the whole interaction can be written as
\begin{equation}
  H^{e-e}=\left(\sum_{(k,k',q) \in B}V^d_{k,k',q}+\sum_{(k,k',q) \not
    \in
    B}V^r_{k,k',q}\right):c^{\dagger}_{k+q}c^{\dagger}_{k'-q}c_{k'}c_k:
  . \label{sA8}
\end{equation}
Both Fr\"ohlich's and the flow equation approach yield the same first
term, i.e. both Eqs. $(\ref{sA3})$ and $(\ref{sA4})$ yield the
result (\ref{sA7}) for $V_{k,k',q}^d$. Only the second term
$V_{k,k',q}^r$ in (\ref{sA8}) differs. Thus the differences between
the interactions $(\ref{sA3})$ and $(\ref{sA4})$ can be seen as a
different generalization from processes with $(k,k',q) \in B$ to
processes with $(k,k',q) \not \in B$. The accordance in $V^d_{k,k',q}$
grants the independence of the self-energy part of the one-particle
energies.

Now the origin of the difference between the two interactions
$(\ref{sA3})$ and $(\ref{sA4})$ has to be investigated. For that it is
useful to formulate the transformation described by the flow equation
as an $l$-independent transformation. To do so the flow equations
\[ \frac{dH}{dl} =[\eta ,H] \]
will be interpreted as an infinitesimal formulation of the unitary
transformation
\[ H(l) =U(l)H(0)U^{\dagger}(l) ,\]
where
\[ U(l):={\rm T}_l e^{\int_{0}^{l}\eta (l')dl'} .\]
Here $l$-ordering is defined in the same way as time-ordering.\\ Thus
to get rid of the $l$-dependence one can introduce
\[ e^{-S}:=U(\infty).\]
Then for $S$ the following expansion holds
\begin{equation}
  S =S_1+ S_2 +... \equiv - \int_{0}^{\infty}dl \eta
  (l)-\frac{1}{2}\int_{0}^{\infty} dl \int_{0}^{l}dl' [\eta (l),\eta
  (l')] +...\label{com1}
\end{equation}
and the neglected terms are again of order $|M_q(0)|^3$. Here $\eta$
denotes again the choice (\ref{fg0a}). Using the solution (\ref{fg7e})
the first term of the last equation becomes the generator of the
Fr\"ohlich-transformation, while for the second term one obtains
\begin{eqnarray}
  S_2 & = & - \frac{1}{2}\sum_{k,k',q} |M_q(0)|^2 \nonumber \\ & &
  \left\{ \frac{\alpha _{k,q}}{\beta_{k',-q} (\alpha _{k,q}^2+\beta
      _{k',-q}^2)}-\frac{1}{\alpha_{k,q}\beta_{k',-q}}
    -\frac{\beta_{k,q}}{\alpha_{k',-q}(\beta_{k,q}^2+\alpha_{k',-q}^2)}+
\frac{1}{\beta_{k,q}\alpha_{k',-q}}\right
    \} \nonumber \\ & & \times c^{\dagger}_{k+q} c^{ }_k
    c^{\dagger}_{k'-q}c^{ }_{k'-q}. \nonumber\\ & & + \mbox{ terms of
      the structure }a^{\dagger}a c^{\dagger}c.
\end{eqnarray}
To derive the result (\ref{sA3}) by using this modified
Fr\"ohlich-transformation 
up to order $|M_q(0)|^2$ only $[\sum_{k} \varepsilon _k
c^{\dagger}_kc_k,S_2]$ has to be taken into account. Again
the generated interaction of the form (\ref{tH1}) will be 
neglected\footnote{Actually
  these terms together with $-\int_{0}^{\infty}dl \eta^{(2)}(l)$
  guarantee that $H(\infty)$ does not contain any interactions of the
  form (\ref{tH2}).}.
It is easily verified that 
\begin{equation}
  [\sum_{k} \varepsilon _k c^{\dagger}_kc_k,S_2] = \sum_{k,k',q}
  (V_{k,k',q} (\infty) -V_{k,k',q}^F ) c^{\dagger}_{k+q}
  c^{\dagger}_{k'-q}c_{k'}c_k
\end{equation}
holds.

Thus the change in the effective electron-electron interaction is
caused by a change of Fr\"ohlich's generator. This modification is
produced by carrying out the $l$-ordering. Thus it is a consequence of
the $l$-dependence of $\eta$ respectively of the $l$-dependence of the
coefficients of $H$. Or more rigorously spoken it is the flow of the
couplings which changes the unitary transformation.

By comparing $(\ref{sA3})$ and $(\ref{sA4})$ it is obvious that the
flow equations yield a result which is less singular, i.e.
$V_{k,k',q}$ diverges\footnote{We will return to this point in Section
  \ref{pot2}.} only for the special case $\alpha_{k,q}=\beta
_{k',-q}=0$. Nevertheless it is not clear which transformation is
better with respect to the neglected interactions. Although the
modification $S_2$ given above eliminates most of the singularities of
Fr\"ohlich's result it is not clear by now if this modification $S_2$
is the best one can choose. By changing the coefficient of $S_2$
independent of any $\eta$ a wide variety of possible corrections to
$V^F_{k,k',q}$ can be achieved. It is still an open question what the
side-effect of these modifications will be. But one should be aware
that modifications of $S_2$ are equivalent to modifications of the 
the unitary transformation. Only for real processes (where energy is 
conserved), one does not have a choice to modify the effective interaction 
in this perturbative approach.

\section{The Asymptotic Behavior}
\label{sas}

By now the elimination of the electron-phonon coupling has been considered 
up to second order. A main advantage of the flow-equations is that in 
certain cases one can go beyond perturbation theory. Then the continuous 
change of the single particle energies has the interesting effect, that for 
processes which are real for a certain value of the flow parameter $l$, this 
energy-conservation does no longer hold for other values of $l$. As Kehrein, 
Mielke and Neu \cite{Keh2} have shown for the spin-boson-problem this allows a 
complete elimination of the coupling even if energy-conservation holds 
asymptotically, i.e. for $l$ approaching infinity.
The renormalization of the electron-energies 
$\varepsilon _k$ will be neglected, assuming that the electron-phonon
coupling is not strong enough to cause a significant change 
of $\varepsilon _k$. This holds at least for weak electron-phonon
coupling\footnote{In deriving Eqs. $(\ref{fg5})$ and (\ref{fg6})
  this assumption already has been used. There the $l$-dependence of
  the fermi-functions $n_k$ has be neglected. This can be done for
  small differences $V_{k,k',q}(\infty)-V_{k,k',q}(0)$. This
  assumption is consistent with the BCS-theory, where one neglects the
  self-energy.}. But special cases like heavy fermions are excluded.
Furthermore it is assumed that the influence of the lines two to five
in Eq. (\ref{fg1}) is negligible\footnote{It can be seen by a
  self-consistent analysis that this assumption does not modify the
  results of this section, i.e. the asymptotic behavior of
  $\omega_q$.}. Thus the dominating part of $M_{k,q}$ becomes
\begin{equation}
  M_{k,q}(l) = M_q(0) e^{- \int_{0}^{l}dl'\left ( \varepsilon _{k+q}
    -\varepsilon _k+ \omega _q (l')\right )^2} \label{fg7}
\end{equation}
It should be emphasized that this ansatz corresponds to a highly
non-perturbative behavior. By taking into account the change of
$\omega _q$ in the order of $|M_q(0)|^2$ the Eq. $(\ref{fg7})$
and hence the Eqs. $(\ref{fg3}) -(\ref{fg5})$ contain terms of
any order of $|M_q(0)|^2$. This corresponds in the sense of
perturbation theory to a summation over a particular subclass of
diagrams.

With these approximations only the behavior of $\omega _q$ for large
$l$ has to be examined. The key to the asymptotic regime lies in the
interpretation of Eq. (\ref{fg4}). The coupling $M _{k,q}$ decays
exponentially as long as $\alpha _{k,q}$ does not lie in the vicinity
of a resonance $\alpha_{k,q} \approx 0$. Thus the behavior of $\omega
_q$ for large $l$ is determined only by the contributions coming from
small $\alpha _{k,q}$.

To develop a more formal understanding of this point one replaces in
Eq. (\ref{fg4}) sums by integrals and
introduces the quantity $\Gamma:=4\pi \cdot \frac{V}{(2\pi)^3}$.
%
%
%

Specializing to three spatial dimensions and the case of quadratic
electron-dispersion $\varepsilon_k=\frac{k^2}{2m}$ one gets
\begin{eqnarray}
  \frac{d \omega _q}{dl} & = & -\Gamma \cdot |M_q(0)|^2 \nonumber \\ &
  & \times \left \{ \int_{-k_F}^{k_F} d\sigma (k_F^2-\sigma ^2) \cdot
  (\frac{-\sigma q}{m} +\varepsilon _q+\omega _q)e^{-2
    \int_{0}^{l}dl'(-\sigma q/m+\varepsilon _q+\omega _q)^2} \right.
  \nonumber \\ & & \quad \left.  +\int_{-k_F}^{k_F} d\sigma
  (k_F^2-\sigma ^2) \cdot (\frac{\sigma q}{m} +\varepsilon _q-\omega
  _q)e^{-2 \int_{0}^{l}dl'(\sigma q/m+\varepsilon _q-\omega _q)^2}
\right \} , \label{fgw1}
\end{eqnarray}
where $k_F$ stands for the fermi-momentum.
By introducing
$\overline{\omega}_q:= \frac{1}{l} \int_{0}^{l} dl' \omega _q(l') $ 
one can use
\begin{eqnarray*}
& & \exp \left (-2
    \int_{0}^{l}dl'(-\sigma q/m+\varepsilon _q \pm \omega _q)^2
  \right)= \\
& &
\exp \left(-2\int_{0}^{l} \omega _q ^2(l') dl'
  +\frac{2}{l}\left(\int_{0}^{l}\omega _q(l')dl' \right )^2 \right)
\cdot \exp \left(-2(-\sigma q/m+\varepsilon _q \pm \overline{\omega}
_q)^2 \cdot l \right) .
\end{eqnarray*}
This allows to express the integral over $\sigma $ in terms of
elementary functions and the error integral.
For large $l$ and for values of $q$ which are not too large 
the  integrals can be extended from $-\infty$ to $+\infty$.
By neglecting the 
exponentially decaying terms the result is
\begin{eqnarray}
  \frac{d \omega _q}{dl} & = & -2 \cdot \Gamma \cdot |M_q(0)|^2 \exp
  \left(-2\int_{0}^{l} \omega _q ^2(l') dl'
    +\frac{2}{l}\left(\int_{0}^{l}\omega _q(l')dl' \right )^2 \right)
    \nonumber \\ & & \times \frac{m^2}{q}\sqrt{\frac{\pi}{2l}} \cdot
    [\overline{\omega} _q(\overline{\omega} _q-\omega
    _q)+\frac{1}{4l}]. \label{asc5a}
\end{eqnarray}
It is worth mentioning that in leading order and besides a constant
Eq. $(\ref{asc5a})$ is identical with the integro-differential
equation describing the asymptotic behavior of the flow equations of
the spin-boson problem \cite{Keh2}. Following the approach of Kehrein,
Mielke and Neu one can assume an asymptotic expansion of $\omega _q$,
i.e. $\omega _q$ decays algebraically
\begin{equation}
  \omega _q(l) \approx \omega _q(\infty) +c_q \cdot l^{- \nu }, \quad
  \nu >0,\quad l\gg 1 , \label{asc5b}
\end{equation}
where $c_q$ is independent of $l$. Using $\overline{\omega} _q-\omega
_q \sim l^{-\nu }$ the integrals in Eq. (\ref{asc5a}) are easily
performed.  The only consistent solution is the case with $\nu =1/2$
and $c_q=\pm 1/2$. Numerical studies suggest $\omega _q' \leq 0$, thus
the ansatz (\ref{asc5b}) yields
\begin{equation}
  \omega _q(l) \approx \omega _q(\infty) +\frac{1}{2 \sqrt{l}} , \quad
  l\gg 1. \label{asc7}
\end{equation}
This solution implies that the approach to the $q$-dependent limit is
independent of $q$. We note however that $\omega_q(l)$ can be rescaled
$\omega _q(l)=q \cdot c_s(lq^2)$ and $M_q(0)=\sqrt{q}M(0)$. Then
Eq. (\ref{asc5a}) yields
\begin{eqnarray*}
  \frac{dc_s(z)}{dz} & = &-2 \Gamma \cdot |M(0)|^2 \cdot \exp(-2
  \int_{0}^{z}c_s^2(z')dz'+2z \cdot \overline{c}_s^2) \\ & & \times m^2
  \sqrt{\frac{\pi}{2z}}\cdot[\overline{c}_s(\overline{c}_s-c_s)+\frac{1}{4z}]
\end{eqnarray*}
with the asymptotic behavior
\[c_s(z) \approx  c_s(\infty)+\frac{1}{2 \sqrt{z}}.\]
Assuming that this asymptotic behavior holds for $z \gg z_1$ then
Eq. (\ref{asc7}) holds for $l \gg l_1=z_1/q^2$. Thus for small
$q$ it is only reached for very large $l$.

To get a better understanding of the behaviour in the asymptotic
regime a new approach has to be chosen. Therefore one has to go back
to Eq. (\ref{asc5a}). If $\omega _q$ decays weaker than $l^{-1}$
than only the term $ \sim
\omega_q(\infty)(\overline{\omega}_q-\omega_q)$ has to be considered.
Hence the other terms will be neglected. An a posteriori
justification of this approximation will be given later. The remaining
integro-differential equation can be transformed into a nonlinear
differential equation. One simply differentiates the first term of
(\ref{asc5a}) and expresses the new generated functions in terms of
$l,\omega _q,\omega _q'$. Introducing the function
\[ v(l):= \int_{0}^{l} \omega _q(l')dl' -l \cdot \omega _q (l) , \] 
the result of this procedure becomes
\begin{equation}
  l^2 \left (v'' \cdot v- v'^2\right )+\frac{1}{2}l \cdot v\cdot
  v'+2v^3 \cdot v' =0 \label{asc10} ,
\end{equation}
where $v'=\frac{dv}{dl}$. Thus the integro-differential equation
(\ref{asc5a}) has been replaced by a much more handable nonlinear
differential-equation.

To make a first connection to the ansatz $(\ref{asc5b})$ one sees that
$(\ref{asc7})$ corresponds to $ v= \frac{\sqrt{l}}{2}$. Thus the
algebraic decay given above is a singular\footnote{singular has to be
  understood as being unable to fulfill arbitrary initial
  conditions.} solution of the differential equation (\ref{asc10}).

In order to solve the differential equation $(\ref{asc10})$ one can
make use of the fact that this equation is equidimensional. Thus the
ansatz
\begin{equation}
  v(l)=\sqrt{l} \cdot \eta (\xi), \quad \xi=\log l/l_0 \label{asc11}
\end{equation}
yields the equation
\begin{equation}
  \eta \cdot \eta '' -\frac{1}{2}\eta \eta '-\eta '^2 +2 \eta ^3\eta'
  +\eta ^4 - \frac{1}{4}\eta ^2 =0 \label{asc12},
\end{equation}
which only implicitly depends on $l$. Here is $\eta '=\frac{d \eta}{d
  \xi}$. As a next step one can substitute
\begin{equation}
  \eta '=-\eta \gamma (x),\quad x=\eta ^2 -\frac{1}{2}\log |\eta|
  \label{asc12b},
\end{equation}
which yields for $\gamma \neq 0$
\begin{equation}
  \gamma '=1-\frac{1}{2\gamma}, \label{asc12c}
\end{equation}
with $\gamma '=\frac{d \gamma}{dx}$. Interpreting this as a
differential equation for the inverse function one gets for $\gamma
\neq 1/2$
\begin{equation}
  \frac{dx}{d\gamma}=\frac{2\gamma}{2\gamma -1} \label{asc13}.
\end{equation}
By integrating the last equation one obtains
\begin{equation}
  x=\gamma+\frac{1}{2} \log |2\gamma -1| +C,\quad C:=const.
  \label{asc14}
\end{equation}
Thus the applied transformations yield a first integral of the
differential equation (\ref{asc10}). Although the remaining Eq.
$(\ref{asc12b})$ can only be solved in special cases it is possible to
get an analytical understanding of the behavior of $\omega _q$ in the
asymptotic regime. One only has to use the fact that $\gamma$ obeys
the autonomous equation $(\ref{asc12c})$ and that the first integral
(\ref{asc14}) is known. The further discussion is organized as
follows. First the solution $\gamma$ will be classified by phase space
arguments. The properties of $\gamma$ determine the behavior of
$\eta$. Thus the behavior of the general solution will become clear
and a connection with the singular solution will be made. Finally the
remaining Eq. (\ref{asc12b}) will be solved for some special
cases.

If the function $\gamma: D \rightarrow R$ is a solution of a first
order autonomous equation then it has exactly one of the following
properties:
\begin{enumerate}
\item[(i)] $\gamma $ is constant and $D=R$.
\item[(ii)] $\gamma $ is injectiv and regular, i.e. $\gamma '\neq 0$
  on $D$.
\item[(iii)] $\gamma $ is regular periodic, i.e. $D=R$ and $\gamma '
  \neq 0$ on $D$.
\end{enumerate}
For a proof of this Theorem see for example \cite{Bro}.

While the singular solution (\ref{asc7}) belongs to type (i) the
solution for arbitrary initial conditions has to be classified now.
For that the function $x$ will be interpreted as function of $\gamma$, as
shown in Fig. \ref{Abb41}. It follows from the definition
$(\ref{asc12b})$ that $x \geq x_{min}:=x(|\eta| =1/2)$ (see also
Fig. \ref{Abb42}). Thus the value $\eta =1/2$ corresponds to the
minimum of the transformation $x=x(\eta)$. Because there exists a
lower bound $x_{min}$ the image of the function $\gamma =\gamma (x)$
is also bounded. Therefore $\gamma$ cannot be bijectiv\footnote{In the
  following we will sometimes refer to the inverse function of
  $\gamma$, which has to be understood as local inverse function.}.\\ 
For general initial conditions the function $x=x(\gamma)$ will be
periodic, as can be easily seen by phase space arguments. Because of
$x=\eta^2-\log|\eta|/2$ the same holds for $\eta =\eta (\log l/l_0)$.
\\ \bild{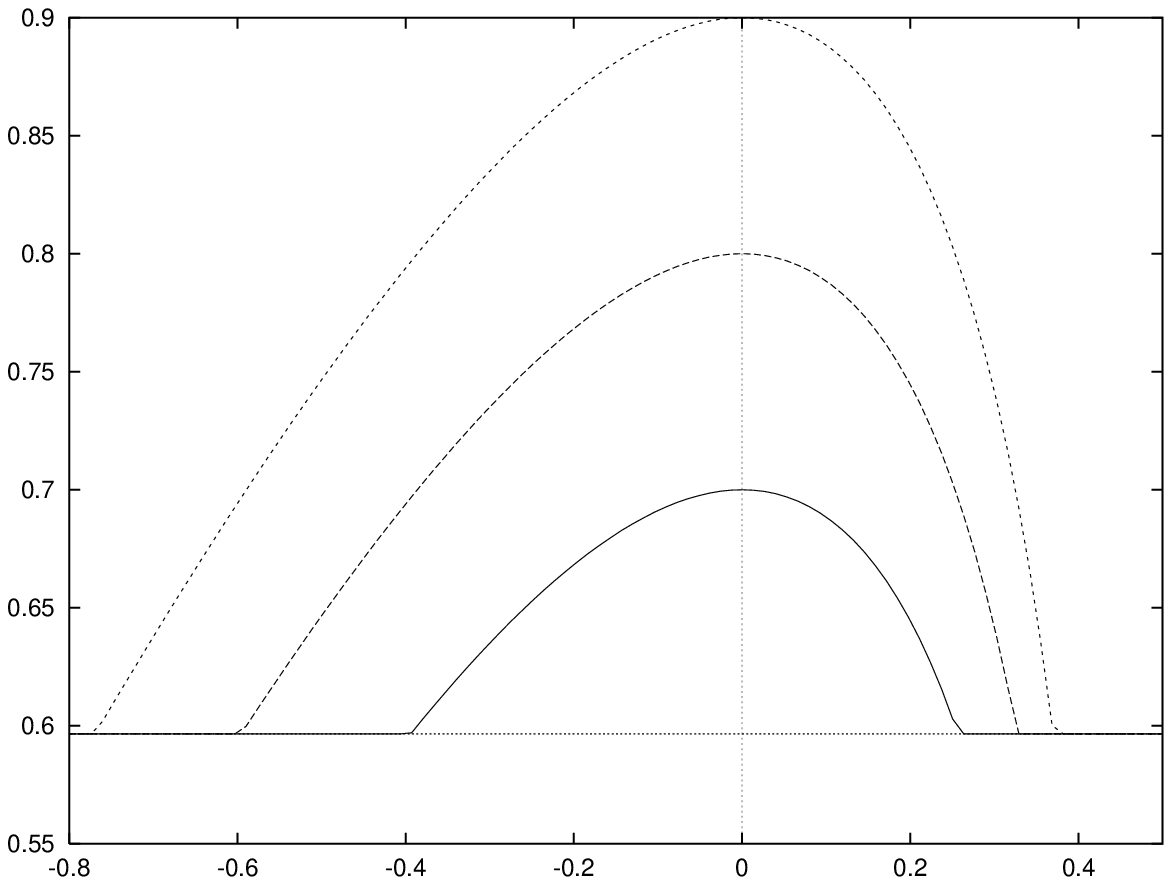}{13.5}{7}{\label{Abb41}$x$ as function of $\gamma$
  to different initial conditions.} Knowing that in general the
function $\eta$ will be periodic some more properties of the solution
can be derived by using Fig. \ref{Abb41}. First it should be
mentioned that the singularities of the differential-equation
$(\ref{asc13})$ split the domain of $x(\gamma)$ in two sections. The
assumption $\omega _q'<0$ implies $\gamma < 1/2$, as can be easily see
{from} Eq. $(\ref{asc11})$. Thus $\gamma =1/2$ is the upper bound
of the domain of $x(\gamma)$. \\ 
\bild{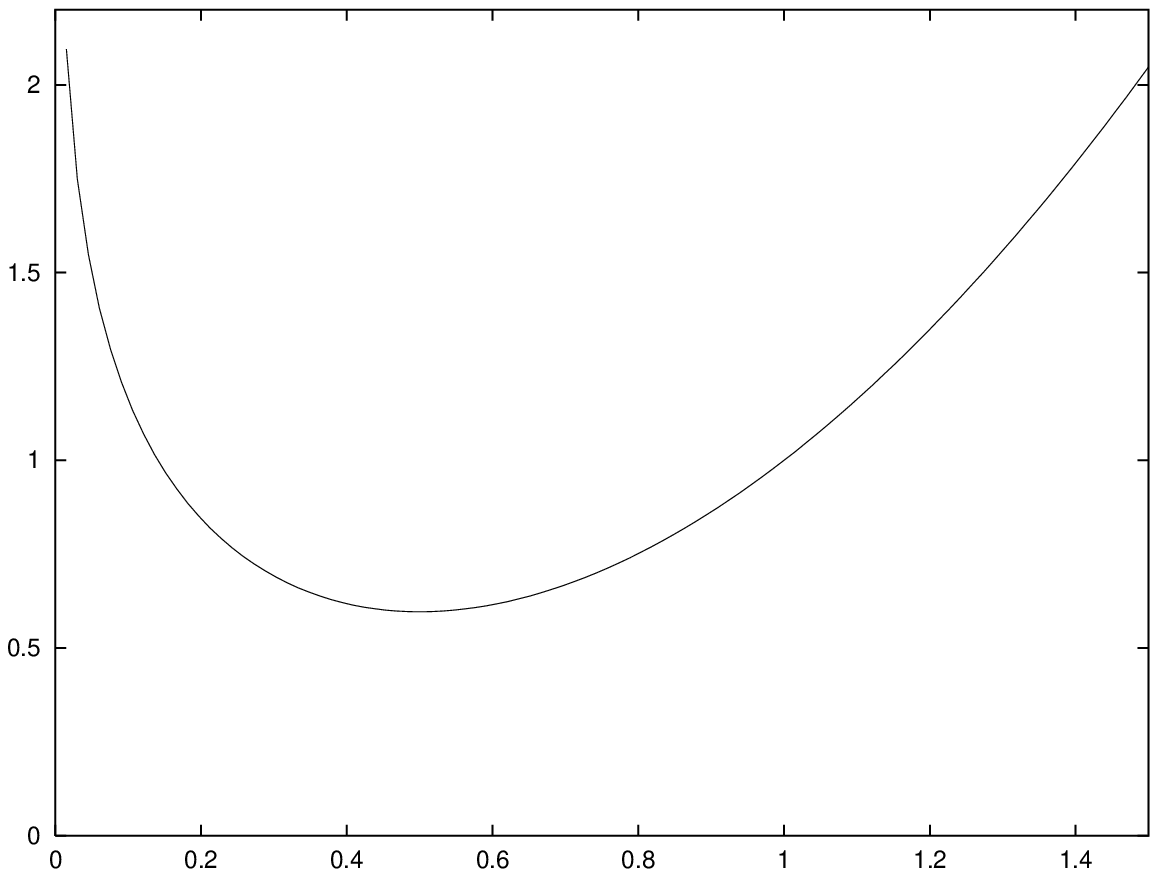}{13.5}{7}{\label{Abb42}$x$ as function of $\eta$. }
Furthermore the point $(\gamma =0,x=x_{min})$ in the phase space
corresponds to a stable trajectory. This follows from the Eqs.
$(\ref{asc12})$ and $(\ref{asc12b})$, which imply $\eta ^{[n]}=0,
\quad \forall n$. Therefore $\eta=const.$, which means that also $x$
and $\gamma$ are constant. This stable trajectory is the singular
solution $(\ref{asc7})$. \\ If general initial conditions $x(\gamma=0)
> x_{min}$ have to be fulfilled the relation $x'(\gamma =0)=0$ still
holds. But the higher derivatives of $x$ no longer vanish. Therefore a
small deviation from the initial condition $x(\gamma =0)=x_{min}$
causes the solution $\gamma =\gamma (x)$ to change from the behavior
of type (i) to type (iii). The trajectory in the phase space is rather
a curve than a point. This curve connects the two turning-points
$\gamma _l,\gamma _r \in \{\gamma:x(\gamma)=x_{min} \}$. It
corresponds to an oscillation of $x$ around $x_{min}$, i.e. of $\eta$
around $1/2$. For the turning-points the relation
\[\left.  \frac{dx}{d \gamma} \right |_{\gamma_l,\gamma_r} \neq 0
\Longleftrightarrow \left .  \frac{d\gamma}{dx} \right |_{x_l,x_r=x_{min}}
\neq 0 \]
holds. This implies that an oscillating trajectory cannot turn into a
constant one. \\ To characterize the oscillation the following
consideration is useful. In Fig. \ref{Abb41} a trajectory $x$ to the
initial conditions $x(\gamma=0)=C_1$ can be seen. This phase space
curve corresponds to an oscillation of $\eta$ along the curve
$x(\eta)$ in Fig. \ref{Abb42}. Therefore the turning- points $\gamma
_l,\gamma _r$ in the phase space correspond to the minimum at $\eta
=1/2$. While the point $x(\gamma=0)=C_1$ represents the greatest
elongation.\\ Defining $\eta _{\pm}:=x^{-1}(C_1)$, where $+$ denotes
the right branch and $-$ left left branch of $x(\eta)$ the following
relation holds:
\[ \eta _+-\frac{1}{2} > \frac{1}{2} - \eta _- . \]
Thus the amplitudes for the elongations $\eta > 1/2$ are greater than
those for $\eta < 1/2$. On the other hand the 'time' for which the
solution stays in the regime with $\eta < 1/2$ is longer than that for
$\eta > 1/2$. This becomes clear by $|\eta '|=\eta \cdot |\gamma|$ and
the fact that the image of $\gamma$ is independent of the branch. But
$|\eta|$ is smaller in the left branch than in the right one.
Therefore $|\eta '|$ is smaller in the left branch.\\ Altogether the
deformation of the oscillations in the Figs. \ref{Abb43}-\ref{Abb46}
can be understood by now.\\ 
\bild{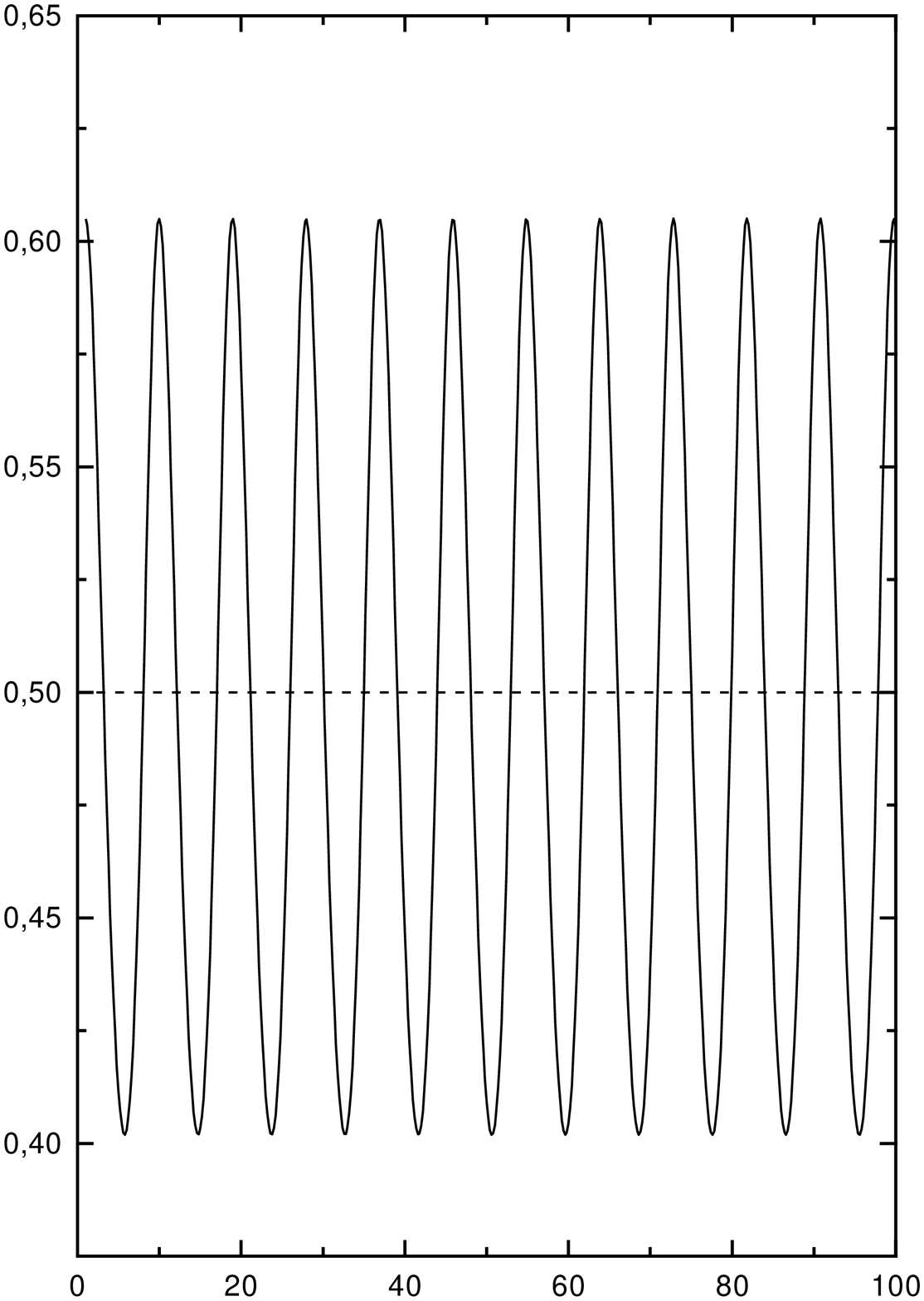}{10}{6}{\label{Abb43}$\eta$ as function of $\log l/l_0$ to
the initial conditions $\eta (\xi _1)=0. 6$, $\eta '(\xi _1)=0$.}
\bild{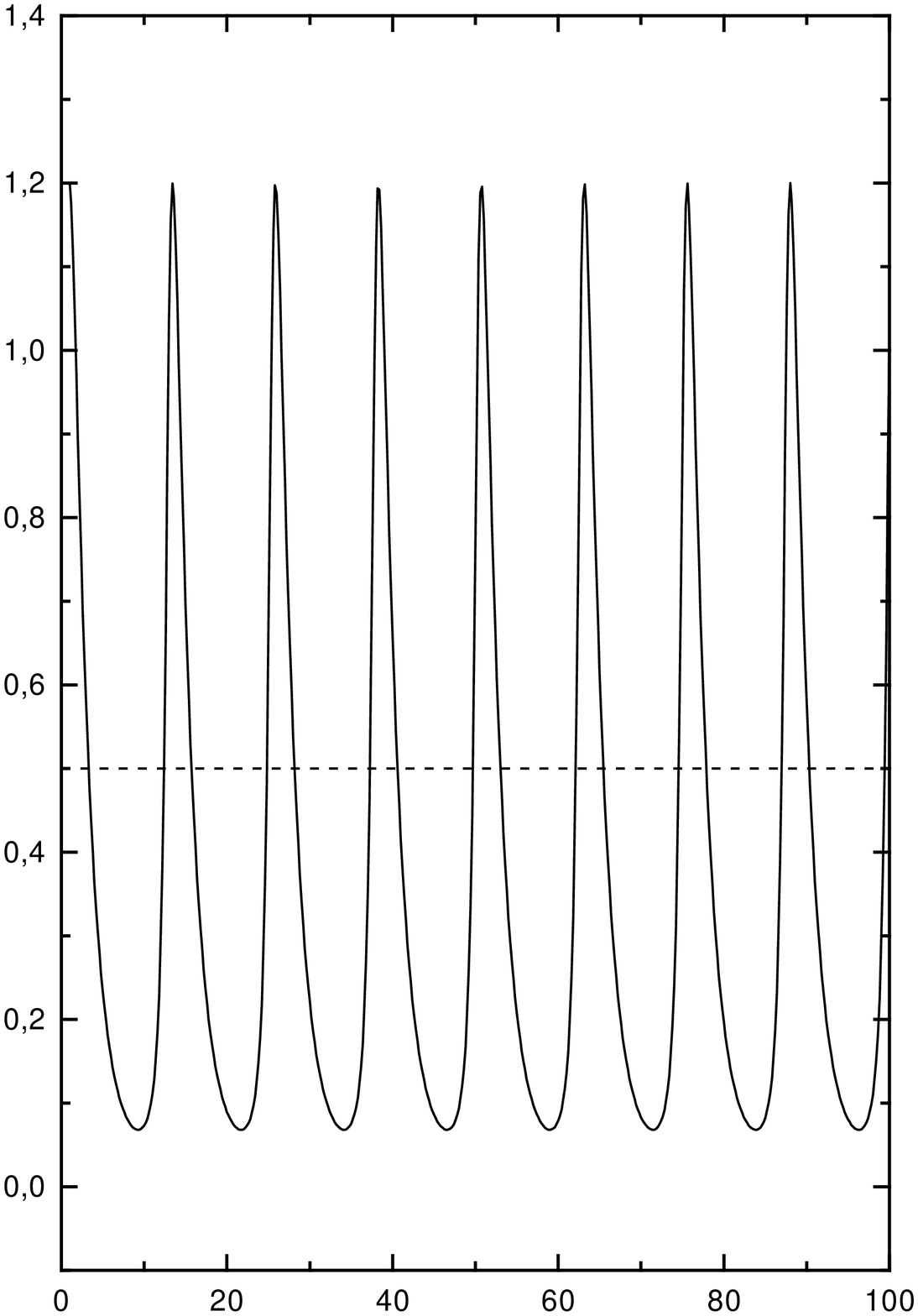}{10}{6}{\label{Abb45}$\eta$ as function of $\log
  l/l_0$ to the initial conditions $\eta (\xi _1)=1. 2$, $\eta '(\xi
  _1)=0$.} \bild{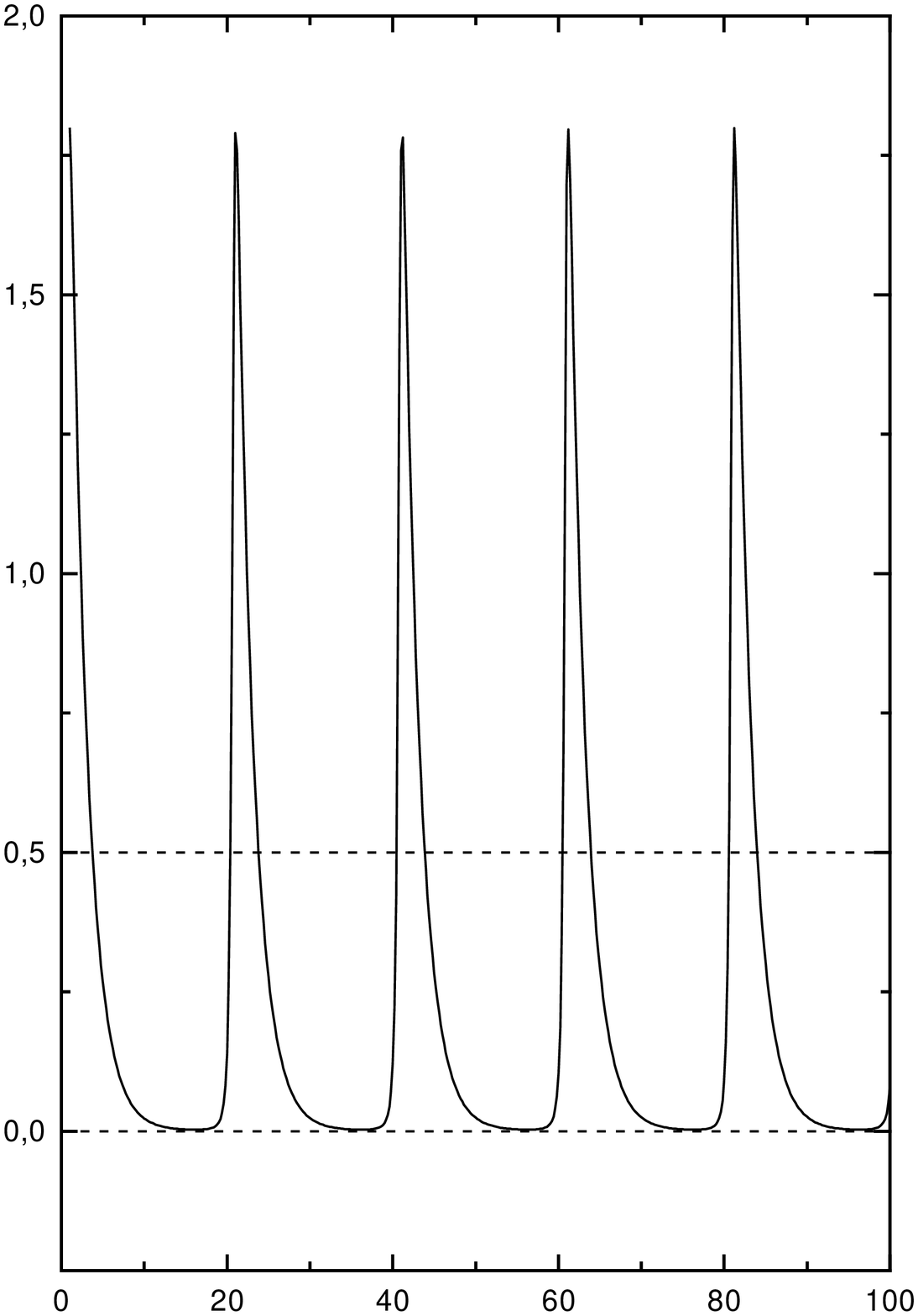}{10}{6}{\label{Abb46}$\eta$ as function of
  $\log l/l_0$ to the initial conditions $\eta (\xi _1)=1. 8$, $\eta
  '(\xi _1)=0$. The sequence of figures corresponds to a proceeding
  deviation from the singular solution.}

The above mentioned differences in amplitude and 'remain-time' in the
cases $\eta >1/2$ and $\eta < 1/2$ suggest that the relation $\langle
\eta ^2 -1/4 \rangle \equiv \overline{\eta ^2}-1/4=0$ holds. Here
$\langle ...  \rangle$ denotes an average over one period. In analogy
to the prove of the virial theorem one can try to find a periodic
function $\tilde{f}(\eta,\eta ')$ fulfilling the relation
\begin{equation}
  \frac{d\tilde{f}}{d \xi}=\eta ^2 -\frac{1}{4} .\label{vsb1}
\end{equation}
The condition $(\ref{vsb1})$ becomes easier to handle if again a
change to the variables $x$ and $\gamma$ is made. Letting
$f(x,\gamma)=\tilde{f}(\eta,\eta ')$, one has to fulfill the condition
\begin{equation}
  \frac{\partial f}{\partial \gamma} \cdot \left
  (1-\frac{1}{2\gamma}\right)+\frac{\partial f}{\partial x}=-
  \frac{1}{2\gamma}. \label{vsb2}
\end{equation}
It is sufficient to take only analytical $f$ into account:
\begin{equation}
  f(x,\gamma) =f_0(x)+f_1(x) \cdot \gamma+ \frac{1}{2!}f_2(x) \cdot
  \gamma ^2 +... \label{vsb3}
\end{equation}
Therefore one obtains the recursional relations
\begin{eqnarray}
  f_1 & = & 1 \\ nf_n-\frac{1}{2}f_{n+1}+n\frac{df_{n-1}}{dx} & = & 0
  \mbox{ for } n \geq 1 . \label{vsb4}
\end{eqnarray}
The function $f_0(x)$ can be chosen arbitrarily. The ansatz $f_0(x)
\sim x^m,$ $m \geq 2$ implies that $f_n$ is a polynom of degree
$(m-1)$ in $x$ for $n \geq 2$. In order to keep the analysis as simple
as possible only the two following cases will be discussed:\\ 
\underline{{\bf 1.case:}}$f_0(x)=C=const.$: It follows easily by
induction that $f_{n+1}(x)=2^{n}\cdot n!$ holds for $n \geq 0$.
Together with Eq. $(\ref{vsb3})$ this yields
\begin{equation}
  f(x,\gamma)=C-\frac{1}{2}\log \left(1-2 \gamma \right) .\label{vsb5}
\end{equation}
Because of $\gamma < 1/2$ the last equation is welldefined.\\ 
\underline{{\bf 2.case:}}$f_0(x)=-x$: Here $f_n(x) \equiv 0$ holds for
$n \geq 2$, and therefore
\begin{equation}
  f(x,\gamma)=\gamma -x . \label{vsb6}
\end{equation}
Because of Eq. (\ref{asc14}) the Eqs. (\ref{vsb5}) and
(\ref{vsb6}) are actually identical. Thus the oscillating behavior of
the function $\eta$ is completely determined by the first integral
(\ref{asc14}). Simultaneously it is shown that $\langle
\eta ^2  \rangle =1/4$.

{From} the definition of $v$ and Eq. ($\ref{asc11}$) one obtains
\[ \frac{d \omega _q}{dl} = -l^{-3/2} \cdot (\frac{1}{2}\eta + \eta ' ). \]
{From} this one can conclude
\begin{equation}
  \omega _q (l)=\omega _q(\infty)+ c_q
  \frac{1}{\sqrt{l}},\quad l \gg 1 . \label{ai1}
\end{equation}
Here $c=c_q$ obeys the differential equation
\begin{equation}
 \frac{c}{2}-c' = \frac{\eta}{2}+\eta ',\quad c'=\frac{dc}{d\xi} \label{az1}
\end{equation}
with the solution
\[c(\xi) = \exp(\xi/2) \int_{\xi}^{\infty} (\frac{\eta}{2}
+\eta') \cdot \exp(-\xi/2)d\xi. \]
Since $\eta$ is periodic in $\xi$, this also holds for $c$. One notes, that
$\eta /2+\eta' = \eta (1/2-\gamma)$ is positive, since both factors on 
the right hand side have this property. Thus $\omega _q$ decays monotonically
as
$l^{-1/2}$ and the term $(4l)^{-1}$ in Eq. (\ref{asc5a}) is asymptotically 
negligible. Thus the approximations leading to Eq. (\ref{asc10}) are 
indeed justified.

For the following it will be of importance to show that the average of $c^2$ 
with respect to $\xi$ equals $1/4$. From the differential equation 
(\ref{az1}) one obtains
\[ c^2-\eta^2 = (c+\eta)(c-\eta) = 2 (c'+\eta')(c+\eta) = 
\frac{d(c+\eta)^2}{d\xi} .\]
Since the right hand side is a total derivative the averages of $c^2$ and
 $\eta ^2$ coincide. Thus $\langle c^2 \rangle = 1/4$.

Finally two special solutions of Eq. (\ref{asc12}) will be
given.\\ 
\underline{{\bf Nearly harmonic case:}} Here it is assumed that
$\eta$ deviates only weakly from the singular solution (corresponding
to Fig. \ref{Abb43}). By expanding the deviation of $\eta ^2$ from 
 $1/4$ in powers of the amplitude $A$ with $A \ll 1$ the Eq. 
(\ref{asc12}) yields the solution
\begin{eqnarray*}
  \eta (\xi)^2 & = & \frac{1}{4}+A \cdot \cos
  \left(\frac{1}{\sqrt{2}}y \right ) \\ 
  & & +A^2 \cdot \left [-\frac{\sqrt{2}}{3}\sin\left(\frac{2}{\sqrt{2}}y 
 \right )+\frac{4}{3}\cos\left(\frac{2}{\sqrt{2}}y \right ) \right ]\\ 
& & + A^3 \cdot \left[\frac{5}{4}\cos\left(\frac{3}{\sqrt{2}}y \right
    ) -\sqrt{2}\sin\left(\frac{3}{\sqrt{2}}y \right )\right ]+ {\cal
      O}(A^4),\\ y & = & \varphi+\xi \cdot \{1-A^2 +{\cal O}(A^4)\}
\end{eqnarray*}
Thus stronger deviations from the singular solution correspond to
smaller frequencies. Observe, that the term proportional to $A^2$ does not 
contain a constant. The same applies for the term proportional to $A^4$ 
which is not given here. This was a first indication for us, that the average 
of $\eta ^2$ might be $1/4$ independent of the amplitude A.

Instead of averaging the solution the algebraic decay of $\omega _q$
can be characterized in first order in $A$ by the decay-exponents $\nu
=\frac{1}{2},\frac{1}{2} \pm i \frac{1}{\sqrt{2}}$. Hence harmonic
oscillations correspond to complex decay-exponents.

\underline{{\bf Strongly non-harmonic case:}} Here the solution is
characterized by a strong deviation from the singular solution
(corresponding to Fig. \ref{Abb46}). For the case $\eta \ll 1/2$
Eq. (\ref{asc12}) can be expanded in powers of $\varepsilon$ if
$\eta$ is assumed to be of order $\varepsilon$. In this case the
solution is
\[\eta (\xi)=\varepsilon \exp\{-\frac{1}{2}\xi+C \cdot e^{\xi/2}\}+
{\cal O}(\varepsilon^2), \quad C=const.,\]
where $C$ must be positive to assure that $\eta$ has a positive slope.\\ 
The case $\eta \gg 1/2$ is not so easy to handle. Here $\eta$ can be
expanded at its maximum $M=\eta^2_{{\rm max}}-\frac{1}{2}\log
\eta_{{\rm max}}\approx \eta^2_{{\rm max}}$. In the vicinity of this
maximum $\gamma =\sqrt{M^2-\eta^2}$ holds as can be seen by expanding
Eq. (\ref{asc14}). Thus a good approximation for this regime is
\[\eta (\xi)=\frac{2M e^{-M(\xi -\xi_0)}}{1+ e^{-2M(\xi -\xi_0)}} .\]

\section{Comparison with Fr\"ohlich's Method II: The Degenerate Case}
\label{pot2}
If systems with many electrons and phonons are considered in general
it will be not possible to avoid that the energy difference $\alpha
_{k,q}$ becomes zero. More precisely there will exist electron and
phonon momenta satisfying the relation
\[ \frac{k q \cos \theta}{m} \pm v_s q=0,\]
where $\theta$ denotes the angle between ${\bf k}$ and ${\bf q}$ and
$v_s$ the velocity of sound. The Fr\"ohlich-transformation is not
defined for these $k,q$. Thus the corresponding couplings $M_{k,q}$
are not eliminated in this approach. \\ But even in these cases
the elimination described by the flow equations works. To verify this
the asymptotic behavior of $\omega_q$ has to be taken into account. \\ 
For $\alpha_{k,q}$ the following asymptotic expansion holds:
\[ \alpha _{k,q} (l) \approx \alpha _{k,q}(\infty) + c_q \cdot 
\frac{1}{\sqrt{l}} ,\quad l \gg 1  .  \]
Thus for $\alpha _{k,q}(\infty)=0$ one obtains
\begin{equation}
  M_{k,q} \approx C \cdot \left(\frac{1}{l}\right)^{\overline{c}_q^2}
  =C \cdot \left(\frac{1}{l}\right)^{1/4} ,
\end{equation}
with an appropriate constant $C$. In this case the decay is rather 
algebraic than exponential. \\ 
Hence it is not important that $\alpha_{k,q}$ becomes zero for some
special $l$. It is only important that it is not zero for all $l$.
Thus the elimination of $M_{k,q}$ is a direct consequence of the
renormalization of $\omega_q$.

A similar argumentation was used by Kehrein, Mielke and Neu
\cite{Keh2} for the spin-boson model. There the authors argued that
the coupling to the bosonic bath always is eliminated because of the
renormalization of the tunneling frequency.

A final remark shall be made on the remaining singularities of the
electron-electron interaction (\ref{sA3}). By taking the
renormalization of $\omega_q$ into account only the special values
$\alpha_{k,q} (\infty)=\beta_{k',-q}(\infty)=0$ are critical. It is
easily seen that in this case
\begin{equation}
  V_{k,k',q}(l)= V_{k,k',q}(l_0)+D \cdot \log l/l_0, \quad D:=const.
\end{equation}
holds for $l \gg 1$. Thus there will remain some singularities in our
approach. But 
as long as these special values are multiplied with functions which
stay finite for $l \rightarrow \infty$ they will have no
influence\footnote{This will typically happen if one calculates
  averages of observables.}.

\section{Conclusions}

It has been shown that it is possible to eliminate the electron-phonon
interaction by using continuous unitary transformations. This
elimination causes the renormalization of the coupling functions of
the Hamiltonian described by the flow equations. This change of the
couplings corresponds to the renormalization of the one-particle
energies and to the generation of an effective electron-electron
interaction. In order to set up these differential equation the
generated new interactions have to be neglected.

By analyzing these flow equations some different approaches can be
chosen. First it is possible to apply the approximation used by
Fr\"ohlich. In this case terms of higher order than $|M(0)|^2$ can be
neglected. Furthermore the possibility of degeneracies has to be
excluded. By doing so the flow equations can be solved exactly. By
comparing the results with Fr\"ohlich's one sees that the induced
electron-electron interactions differ. The origin of this difference
lies in the fact that in our approach all couplings depend on $l$.
Thus by going over to a formulation similar to Fr\"ohlich's, i.e.
carrying out the $l$-ordering of $\eta$, it can be seen that the flow
of the coefficients changes the generator of the unitary
transformation. Our result looks much friendlier than Fr\"ohlich's
interaction. But nevertheless it is not clear by now in which sense
our result is 'better' or even 'the best' one can obtain.

It is worth mentioning that Kehrein and Mielke obtain similar
modifications by eliminating the hybridization term in the single
impurity Anderson model by continuous unitary transformations
\cite{Keh3}. The authors show that their approach generates a
spin-spin interaction which differs from the one obtained by the
famous Schrieffer-Wolff transformation. Their induced interaction is
also less singular. But even more important their result shows the
right high-energy cutoff. Thus at least for this model a physical
criterion exists to decide which interaction is 'better'.

Besides this direct comparison with Fr\"ohlich's results it is
possible to prove that our approach indeed eliminates the interactions
between one phonon and one electron. For this purpose the influence of
the renormalization of $\omega_q$ on the elimination of $M_{k,q}$ has
to be taken into account. By doing so the renormalization of
$\omega_q$ in the asymptotic regime is described by an
integro-differential equation. This equation can be transformed into a
differential-equation which is much easier to analyze. By using its
first integral the behavior of $\omega_q$ for large $l$ can be
characterized. As a consequence of the properties of $\omega_q$ the
electron-phonon coupling is always eliminated even in the case of
degeneracies. Furthermore it is shown that the remaining
singularities in the electron-electron interaction are of a very weak 
nature. 

{\bf Acknowledgment.} The authors would like to thank
S. Kehrein and A. Mielke for many helpful discussions.
\begin{appendix}
  
\end{appendix}
\end{document}